\documentclass[conference]{IEEEtran}
\IEEEoverridecommandlockouts
\usepackage{cite}
\usepackage{amsmath,amssymb,amsfonts}
\usepackage{algorithmic}
\usepackage{graphicx}
\usepackage{textcomp}
\usepackage{xcolor}
\usepackage{hyperref}
\usepackage{tikz}
\hypersetup{hidelinks}
\definecolor{lime}{HTML}{A6CE39}
\DeclareRobustCommand{\orcidicon}{
	\begin{tikzpicture}
	\draw[lime, fill=lime] (0,0) 
	circle [radius=0.16] 
	node[white] {{\fontfamily{qag}\selectfont \tiny ID}};
	\draw[white, fill=white] (-0.0669,0.099) 
	circle [radius=0.007];
	\end{tikzpicture}
	\hspace{-2mm}
}

\def\BibTeX{{\rm B\kern-.05em{\sc i\kern-.025em b}\kern-.08em
    T\kern-.1667em\lower.7ex\hbox{E}\kern-.125emX}}
    
\foreach \x in {A, ..., Z}{\expandafter\xdef\csname orcid\x\endcsname{\noexpand\href{https://orcid.org/\csname orcidauthor\x\endcsname}
			{\noexpand\orcidicon}}
}

\DeclareRobustCommand*{\IEEEauthorrefmark}[1]{%
  \raisebox{0pt}[0pt][0pt]{\textsuperscript{\footnotesize #1}}%
}

\begin{document}
\title{Predicting The Stock Trend Using News Sentiment Analysis and Technical Indicators in Spark }

\author{\IEEEauthorblockN{Taylan Kabbani\IEEEauthorrefmark{1,2}\orcidA{},
Fatih Enes Usta\IEEEauthorrefmark{3}\orcidB{}}
\IEEEauthorblockA{\IEEEauthorrefmark{1}Graduate School of Sciences and Engineering, Özyeğin University, Istanbul, Turkey}
\IEEEauthorblockA{\IEEEauthorrefmark{2}Huawei Turkey R\&D Center, Istanbul, Turkey}
\IEEEauthorblockA{\IEEEauthorrefmark{3}Department of Computer Science, Engineering Faculty, Marmara University, Istanbul, Turkey}
}


\maketitle

\begin{abstract}
Predicting the stock market trend has always been challenging since its movement is affected by many factors. Here, we approach the future trend prediction problem as a machine learning classification problem by creating tomorrow\_trend feature as our label to be predicted. Different features are given to help the machine learning model predict the label of a given day; whether it is an uptrend or downtrend, those features are technical indicators generated from the stock's price history. In addition, as financial news plays a vital role in changing the investor's behavior, the overall sentiment score on a given day is created from all news released on that day and added to the model as another feature. Three different machine learning models are tested in Spark (big-data computing platform), Logistic Regression, Random Forest, and Gradient Boosting Machine. Random Forest was the best performing model with a 63.58 \% test accuracy.
\end{abstract}


\section{Introduction}
\label{sec:introduction}
Financial markets are at the heart of the modern economy; they provide an avenue for selling and purchasing assets such as bonds, stocks, foreign exchange, and derivatives. Nevertheless, to profit from such markets, their complicated environment should be well studied. The stock market, in particular, has drawn the attention of many researchers to predict the price trend. The environment of the stock market, which is described as highly volatile with many external factors, have a direct effect on the stock price (historical stock prices, supply, and demand, news releases, etc.), has made the price prediction process very difficult for an individual to perform because it relies on this large amount of data produced by the stock market. Despite some theories \textcolor{blue}{The Efficient Market Hypothesis (EMH[1]} and \textcolor{blue}{The Random Walk theory[2]} that say market movement can not be predicted, many recent studies showed that with the help of machine learning algorithms and big data computing platforms, the trend direction indeed could be predicted. Mainly there are two methods for forecasting market trends. The first method is Technical Analysis which considers the stock's price and volume as the only inputs in forecasting the future trend of the stock. The second method, Fundamental Analysis, studies everything from the overall economy and industry conditions to news releases. In this paper, both methods will be followed to enhance the prediction accuracy of the future stock's trend\textcolor{blue}{[3]}. We use a machine learning algorithm in big data computing platform-Spark to conduct a binary classification prediction (Uptrend, Downtrend) according to stock's daily price information, volume, and other technical indicators. In addition, natural language processing(NLP) was used to analyze financial news data on a given day and give a sentiment score that reflects the polarity of the news (positive or negative) over that day, to be added as a feature to the classification model.

\section{Related Work}
Many research papers have been published in the last decade attempting to achieve high market prediction accuracy, using different methodologies to accomplish the task. One of the latest research papers in 2020 was \textcolor{blue}{Khan(2020)[4]}, where different machine learning algorithms were used on social media, news, and financial stock data to predict the stock's future trend after ten subsequent days, they concluded that combining sentiment analysis of social media and financial news led to decreasing the highest accuracy but increased the overall accuracy of most of the classifiers after day 3, they also reported consistent result using Random Forest algorithm with 77.10\% average accuracy on 10-fold cross-validation. \textcolor{blue}{Kalyani(2016)[5]} also proposed a method to predict the future trend of the stock. However, only by using news sentiment classification, news articles for three years(2013-2016) were retrieved and reprocessed to be fed to three different classification algorithms, Random Forest, Supporting Vector Machine, and Naive Bayes algorithms. RF was the best performing algorithm with testing accuracy ranging from 88\% to 92\%. \textcolor{blue}{Xianya(2019)[6]} implements binary classification prediction of stock trend (rise or fall) in Spark platform to reduce the run time of the models. They reported that when the number of nodes in the cluster increases by 3, the model run time is reduced to two-fifth, indicating the usefulness of using a big data platform such as Spark in machine learning tasks. The classification prediction in this paper is based only on the price information of stocks without sentiment analysis of news; the best performing model was the Random Forest algorithm with a 0.689 AUC score. In another paper \textcolor{blue}{Liu(2016)[7]} machine learning algorithms are used to forecast the future trend of the S\&P 500 index, where different financial features data from 2004 to 2014 was collected. These features were global market indexes, currency rates, commodity price, and technical indicators of S\&P 500 itself (like momentum and rate-of-change). Instead of using the daily stock price, they used the daily return to label the S\&P index upward or downward. Their best performing model was SVM Radial Basis with Function(RBF) kernel by 62.51\% accuracy for the future market trend of the S\&P 500 index. By calculating several technical indicators and using them as additional features in the prediction model \textcolor{blue}{Raşo(2019)[8]} could predict the future price of the BIST-index. More than ten different technical indicators were calculated from daily price information to be fed later to a deep learning algorithm; this paper demonstrates the significance of technical indicators and technical analysis in general when predicting the stock market.

\section{proposed Method}
\label{sec:guidelines}
This section explains our six-step proposed method to predict the next-day trend based on fundamental and technical analysis features. Mainly, the data of three stocks from the US stock exchange will be used to train the machine learning models, Apple Inc. (AAPL), Amazon.com Inc.(AMZN), and Netflix Inc.(NFLX)\\
\textcolor{blue}{Pyspark[13]}, \textit{the Python API written in python to support Apache Spark} is used to implement the project, which significantly reduced the running time of training ML models and allowed us to better tune the hyperparameters by searching a vast grid of model parameters. In addition, using the big data platform in our project enabled us to efficiently deal with the huge news articles we retrieved (9.2 GB). A flow chart illustrates steps and sub-steps in \textcolor{blue}{Fig 1}.

\begin{figure}[!t]
\centerline{\includegraphics[width=\columnwidth]{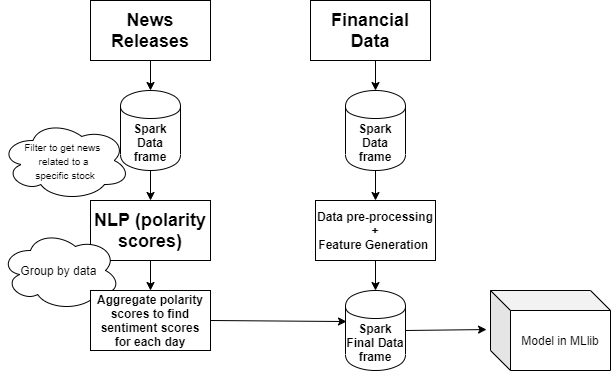}}
\caption{Flowchart Diagram of the Proposed Method }
\label{Fig 1}
\end{figure}

\subsection{Data Collection}
\subsubsection{Financial Data}
We imported the daily stock price data of the mentioned companies for four years (from 2016-01-01 to 2020-04-01) using \textcolor{blue}{Yahoo Finance[9]} which provides a python model to get historical stock data. The data consists of 1069 instances and seven features: date, the highest price of the day, the lowest price of the day, open price, close price, volume, and adjacent close price.

\subsubsection{Financial news data}
News articles data was retrieved from a publicly available dataset \textcolor{blue}{Thompson, Andrew (2017)[10]}, containing 2.7 million articles between January 2016 to April 1, 2020. The news is in raw text format and written in the English language, and they have been collected from well-known news publishers such as Reuters, CNN, CNBC, The New York Times, The Hill, Washington Post, and others. The size of the data is 9.2 GB and consists of six features: date, author, title, article, URL, section, and publication.

\subsection{Data Preprocessing}
The Retrieved news articles are in raw format, so to use the data as input to the machine learning algorithms, we need to preprocess it to an appropriate format. We performed the following steps on the raw news data:
\begin{enumerate}
\item The dataset is converted to Spark data frame. Only Date and Article columns were kept.
\item Regular Expressions have been removed from articles text because they do not possess any sentiment.
\item All letters were converted to lower cases.
\item Because the data set contains 2.7 million news articles related to different topics, we created a set of keywords and functions to search for news related to the three stocks we are interested in.
\item Tokenize each article into word vector.
\item Stop words (such as: is, the, are, an, a) are removed.
\end{enumerate}
The downloaded financial data was clean and contained no missing values; hence no preprocessing was needed. The date feature is dropped from the final data set before feeding it to the machine learning algorithms.

\subsection{Feature Generation}
To improve the accuracy of the classification model, we created some additional features and technical indicators out of the stock's price information using the Window function provided by \textcolor{blue}{SparkSQL[15]}
\subsubsection{Today\_Trend} We create today's trend feature to reflect the trend of the stock on a given day. We generate this feature by subtracting the open price from the close price of the trading day. Based on the returned value, the stock will be classified as Uptrend or Downtrend for the day\textcolor{blue}{[4]}.
\begin{equation}
  \textbf{Today\_Trend} =
  \begin{cases}
    Uptrend & \text{if $P_{c} - P_{o} \geq 0$} \\
    Downtrend & \text{if $P_{c} - P_{o} < 0$}
  \end{cases}
\end{equation}

\subsubsection{Tomorrow\_Trend} Tomorrow's trend feature is the target feature, which we will try to predict by our ML model. This feature describes the trend of the stock on the following day; by calculating the difference between the closing price of the following day and the closing price of today, we can classify tomorrow's trend as Uptrend or Downtrend:
\begin{equation}
  \textbf{Tomorrow\_Trend} =
  \begin{cases}
    Uptrend & \text{if $P_{tom,c} - P_{today,c} \geq 0$} \\
    Downtrend & \text{if $P_{tom,c} - P_{today,c} < 0$}
  \end{cases}
\end{equation}
\subsubsection{RSI} Relative Strength Index (RSI) one of the most essential momentum indicators that most traders use to identify when the stock is being overbought or oversold over a specified period (here we used 14 trading days), the value of RSI ranges between 0-100 and can be calculated from the following formula: $$ \textbf{RSI} = 100 - [\frac{100}{1+ \frac{Average of 14 day's close Up}{Average of 14 day's close Down}}]$$

\subsubsection{SMA}Simple moving average (SMA) calculates the average of a selected range of prices, usually closing prices, by the number of periods in that range \textcolor{blue}{[3]}. Calculated according to the following equation (n = 14 trading days):
$$ SMA = \frac{P_1 + P_2 +P_3 + ... + P_n}{n}$$

\subsubsection{\%K} Stochastic Oscillator indicator, another momentum indicator like RSI, however, it uses support and resistance levels, can be calculated for 14 trading days period as follows:
$$ \textbf{\%K} = 100 * [\frac{P_{C} - LL_{14}}{HH_{14} - LL_{14}}]$$
where:\\
$P_{C}$: Close price. \\
$LL_{14}$: Lowest low price in the past 14 days.\\
$HH_{14}$: Highest high price in the past 14 days.\\

\subsection{Sentiment Analysis}
Sentiment analysis was performed on the processed news articles by using \textcolor{blue}{Vader sentiment analysis tool[11]} which is integrated into \textit{Natural Language Tool Kit (NLTK)}. The tool takes sentences as input, and by using the lexicon, it outputs four different scores, namely, negative, neutral, positive, and compound, representing the highest score of the three scores. Each news article in our data frame was fed to the sentiment analysis tool, and the compound scores were added as a third column to the news data frame, which consisted of the date column and the article's words vector column.
The calculated sentiment score of each news is between -1 and 1. A sentiment score close to 1 indicates that the news is positive, being closer to -1 means that the news is negative, and if the score is around 0, the news is neutral. The advantage of this library is that it does not ignore the order of the words and reserves the word's context meaning.\\
The news data frame is then grouped by date to calculate the overall sentiment score of a given day by aggregating all sentiment scores of news released on that day. The overall sentiment scores of stock are then added as an additional feature to the machine learning models. When the overall sentiment is high on a given day, the stock tends to receive much attention from traders and investors.
\begin{figure}[!t]
\centerline{\includegraphics[width=\columnwidth]{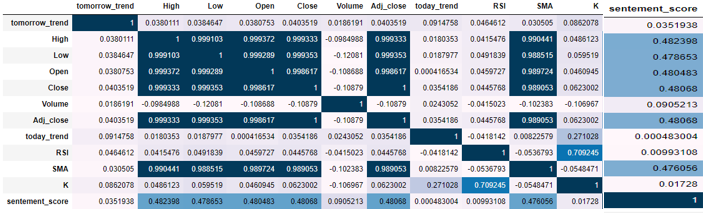}}
\caption{Pair-wise correlation matrix}
\label{Fig 2}
\end{figure}
\subsection{Feature Correlation Analysis}
Here, we show the degree of correlation between our data's final set of features. Pair-wise correlation matrix is created as shown in \textcolor{blue}{Fig 2}. We can infer a high correlation between Close, Open, High, Low, Adj Close, and SMA features from the matrix. Only SMA, High, Close features were kept from these highly correlated features in the final data set.

\subsection{Applying machine learning algorithms}
In this paper three machine learning algorithms are implemented and compared on three different stocks using \textcolor{blue}{MLLib[14]} in Spark, which gives the ability to run machine learning models in parallel. Selected classifiers are \textbf{Logistic Regression}, \textbf{Random Forest} and \textbf{Gradient Boosting Machine}.\\
The final data set of each stock that will be fed to the ML algorithms consisted of 9 features, and 1063 instances see \textcolor{blue}{Fig 3}. Before applying the models, the final data sets are split into 80\% train and 20\% test sets.
\begin{figure}[!t]
\centerline{\includegraphics[width=\columnwidth]{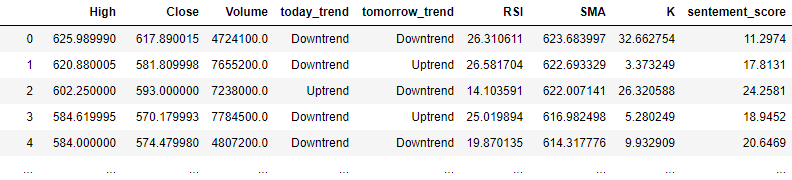}}
\caption{A view of the final data set be fed to ML pipe}
\label{Fig 3}
\end{figure}
\begin{itemize}
\item \textit{ML pipeline:} Like Scikit-learn, MLlib in spark also supports machine learning pipelines to efficiently run several models and perform data preprocessing in sequential steps. In addition, pipelines prevent potential data leakage between test and train data. The ML pipeline in this paper consisted of the following steps:
\begin{itemize}
\item Data Normalization: before applying the classifier, numerical data is normalized to have a range of values between 1 and -1, using MinMaxScaler().
\item One-hot encoding: MLlib provides a function used in R language, RFormula(), which enables the user to separate label variable (tomorrow\_trend) from the rest of the features and automatically converts character features into numerical ones (One-hot encoding).
\item Hyperparameters Optimization: Using the advantage of MLlib in running machine learning tasks in parallel, hence reducing the running time of the models, we tested a wide range of hyperparameters for each model to better tune them and achieve better accuracy \textcolor{blue}{Table 1}.

\begin{figure}[!h]
\centerline{\includegraphics[width=\columnwidth]{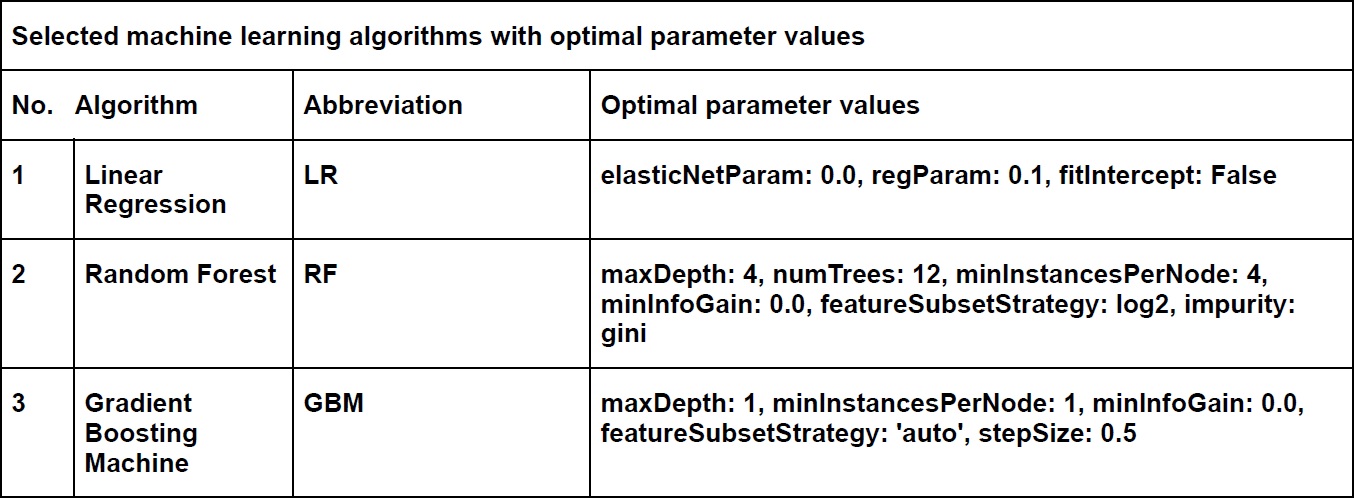}}
\caption{Table 1}
\end{figure}
\item Performance Evaluation: Since our problem is a binary classification problem, selected classifiers' performance will be compared based on four metrics, Accuracy, Recall, Precision, and f1 score. In order to avoid overfitting and select the best parameters for the prediction models, k-fold Cross-Validation(CV) is used to train the models and select the best parameters. The number of folds selected is ten which is recommended by \textcolor{blue}{Kohavi[12]}. The CV overall accuracy estimation is measured by taking the average of ten individual accuracies of the 10-fold for each model on the training phase. The test split is then used to check the test accuracy of each model.
\end{itemize}
\end{itemize}

\section{Results \& discussion}
This section shows the results obtained from predicting the future trend of AAPL, AMZN, and NFLX stocks by the selected models defined in section 3. The evaluation metrics are (1) Accuracy, which represents how often the model is correct (2) Precision, which indicates the percentage of predicted labels that are correctly predicted, (3) Recall, which measures the percentage of how many true labels returned, and (4) F-measure, which
measures the weighted average of precision and recall. Experimental results showed that the best performing model was Random Forest with 0.655 accuracy on the training set and 0.6358 on the testing set. The rest of the models' performances are reported in \textcolor{blue}{Table 2}

\begin{figure}[!h]
\centerline{\includegraphics[width=\columnwidth]{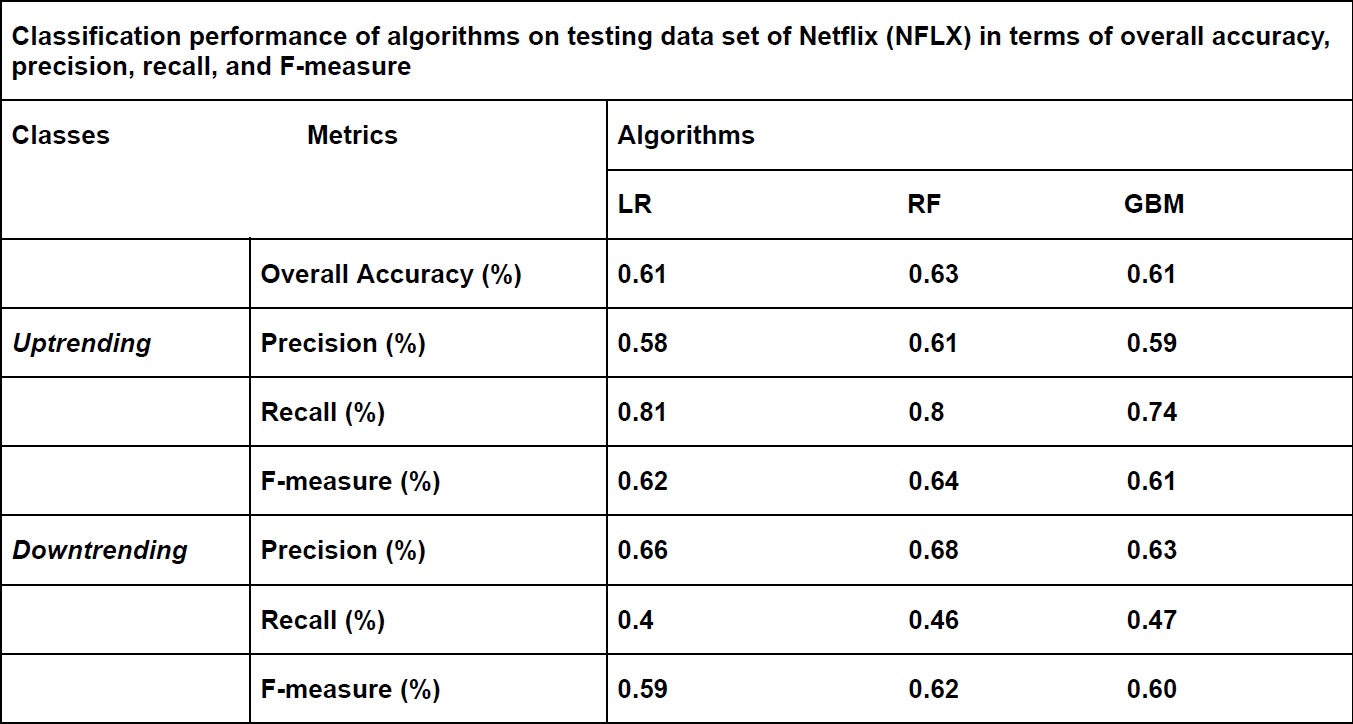}}
\caption{Table 2}
\end{figure}

\section{Conclusion \& future work}
In this paper, the big data platform Spark is used to predict the stock's future trend (uptrend or downtrend) by using technical indicators and news sentiment analysis.
The overall accuracy we obtained is fairly good enough when it comes to predicting the stock market movements since the market trend could be affected by many random factors other than news and price information. However, the shortcoming of the prediction accuracy is due to the insufficient publicly available stock news data sets, so we end up having a small number of instances (only from 2016 to 2020).
In future projects, prediction accuracy can be enhanced by performing time-series analysis, retrieving more instances, adding more features, and using deep learning algorithms that perform better than traditional machine learning algorithms.


\begin{thebibliography}{00}

\bibitem{b1} Malkiel, B. G., \& Fama, E. F. (1970). Efficient Capital Markets: A Review Of Theory And Empirical Work*. \textit{The Journal of Finance}, 25(2), 383–417. doi: 10.1111/j.1540-6261.1970.tb00518.x.
\bibitem{b2} Horne, J. C. V., \& Parker, G. G. (1967). The Random-Walk Theory: An Empirical Test. Financial Analysts Journal, 23(6), 87–92. doi: 10.2469/faj.v23.n6.87
\bibitem{b3} Investopedia. (n.d.). Retrieved from https://www.investopedia.com/
\bibitem{b4} Khan, W., Ghazanfar, M.A., Azam, M.A. et al. Stock market prediction using machine learning classifiers and social media, news. J Ambient Intell Human Comput (2020).
\bibitem{b5} Kalyani, Joshi, H. N., J., \& Rao. (2016, July 7). Stock trend prediction using news sentiment analysis. Retrieved from https://arxiv.org/abs/1607.01958.
\bibitem{b6} Xianya, J., Mo, H., \& Haifeng, L. (2019). Stock Classification Prediction Based on Spark. Procedia Computer Science, 162, 243–250. doi: 10.1016/j.procs.2019.11.281.
\bibitem{b7} Liu, C., Wang, J., Xiao, D., \& Liang, Q. (2016). Forecasting S\&P 500 Stock Index Using Statistical Learning Models. Open Journal of Statistics, 06(06), 1067–1075. doi: 10.4236/ojs.2016.66086.
\bibitem{b8} Raşo, H., \& Demirci, M. (2019). Predicting the Turkish Stock Market BIST 30 Index using Deep Learning. Uluslararası Muhendislik Arastirma Ve Gelistirme Dergisi, 253–265. doi: 10.29137/umagd.425560
\bibitem{b9} Yahoo-finance. (n.d.). Retrieved from https://pypi.org/project/yahoo-finance/
\bibitem{b10} Thompson, Andrew (2017). All the News 2.0 : 2.7 million news articles (2016 -2020).(https://components.one/datasets/all-the-news-2-news-articles-dataset/)
\bibitem{b11} Hutto, C.J. \& Gilbert, E.E. (2014). VADER: A Parsimonious Rule-based Model for Sentiment Analysis of Social Media Text. Eighth International Conference on Weblogs and Social Media (ICWSM-14). Ann Arbor, MI, June 2014
\bibitem{b12} Kohavi R (1995) A study of cross-validation and bootstrap for accuracy estimation and model selection. IJCAI 14(2):1137–1145
\bibitem{b13} Welcome to Spark Python API Docs!¶. (n.d.). Retrieved from https://spark.apache.org/docs/latest/api/python/index.html.
\bibitem{b14} MLlib: Apache Spark. (n.d.). Retrieved from https://spark.apache.org/mllib/
\bibitem{b15} Spark SQL \& DataFrames: Apache Spark. (n.d.). Retrieved from https://spark.apache.org/sql/

\end{thebibliography}
\end{document}